\documentclass[%
 reprint,
groupedaddress,
showpacs,
 amsmath,amssymb,
 aps,
 prd,
 longbibliography,
 lengthcheck,%
]{revtex4-1}

\usepackage{graphicx}
\usepackage{epsfig}
\usepackage{epstopdf}
\usepackage{dcolumn}
\usepackage{bm}
\usepackage{float}

\begin{document}

\title{Highly relativistic spin-gravity-$\Lambda$ coupling}

\author{Roman Plyatsko, Mykola Fenyk, and Volodymyr Panat}

\affiliation{Pidstryhach Institute for Applied Problems in
Mechanics and Mathematics\\ (National Academy of Sciences of Ukraine), \\ 3-b Naukova Street,
Lviv, 79060, Ukraine}

\date{\today}

\begin{abstract}
The effects of  highly relativistic spin-gravity coupling  in the Schwarzschild-de Sitter background  which follow from the Mathisson-Papapetrou equations  are investigated. The dependence of  gravitoelectric and gravitomagnetic components of  gravitational field on the velocity of an observer which is moving in Schwarzschild-de Sitter's background is estimated. The action of  gravitomagnetic components on a fast moving spinning particle is considered. Different cases of the highly relativistic circular orbits of a spinning particle which essentially differ from the corresponding geodesic orbits are described.

\end{abstract}


\maketitle

\section{ Introduction}

According to general relativity, many basic properties of gravity have been discovered through investigations of test particle motions in gravitational fields of different massive sources. The classical results concerning motions of a simple particle which follows geodesic lines in Schwarzschild's and Kerr's metrics are presented, for example, in \cite{Misner, Chandra}. If a test particle possesses inner rotation/spin, in general its motion differs from geodesic. The equations which describe motions of a macroscopic spinning particle in gravitational fields were obtained in \cite{Mathis}
and have been rederived  later in \cite{Papa, Tul, Dix, Taub, Bart} and in many other papers: now these equations are known as {\it Mathisson-Papapetrou} (MP) or {\it Mathisson-Papapetrou-Dixon} (MPD) equations. In addition, it was shown that in a certain sense these equations
follow from the general relativistic Dirac equation as a classical
approximation \cite{Dirac}.

In the first papers where some partial solutions of the MP equations were  investigated  the Schwarzschild \cite{Cor, Mic}, Melvin \cite{Das}, and Lense-Thirring \cite{Pras} metrics were taken into account. The physical effects of the spin-gravity interaction in Kerr's spacetime are presented in \cite{Mash-71, Wald, Rasb, Tod, Pl82} (more full bibliography can be found in \cite{Sem, Pl88} and in other papers). In some recent publications the MP equations are studied for  the Schwarzschild-de Sitter spacetime \cite{Stu, Mort, Kunst, Gwak}.

As usual, the effects of the spin-gravity coupling on spinning particle motion are very small. However, under certain specific conditions, when the orbital velocity of a spinning particle relative to a source of the gravitational field is close to the speed of light, role of the spin-gravity coupling becomes 
much greater. Different types of highly relativistic motions of a spinning particle in Schwarzschild's and Kerr's background were studied in \cite{Pl98, Pl01, Pl05, Pl11, Pl12, Pl15, Pl16b, PlN}. It was shown that depending on the correlation of signs of the spin and the particle's orbital velocity the spin-gravity coupling acts as a significant repulsive or attractive force.

What is the dependence of the influence of the cosmological constant on a spinning particle for its different velocities relative to the Schwarzschild source of the gravitational field? Answering this question is the purpose of this paper. Note that now the role of  the cosmological constant is widely considered in cosmology and astrophysics.

The paper is organized in the following way. Section II is devoted to the consideration of the gravitoelectric and gravitomagnetic components of the single gravitational field in general relativity which can be estimated by an observer moving in  the Schwarzschild-de Sitter background with arbitrary velocity. In Sec. III we present necessary information about the MP equations and then study the spin-gravity coupling in the Schwarzschild-de Sitter background from the point of view of an observer comoving with the spinning particle. Different cases of  highly relativistic circular orbits of a spinning particle in this background which are caused by the strong 
spin-gravity repulsion or attraction are in the focus of Sec. IV.  We conclude in Sec. V.

\section{Gravitoelectric \\ and gravitomagnetic components \\ from the point of view \\ of a moving observer in the \\ Schwarzschild-de Sitter background}

We use the Schwarzschild-de Sitter metric in the standard coordinates
 $x^1=r, \quad x^2=\theta, \quad x^3=\varphi, \quad x^4=t$. 
Then the nonzero components of the metric tensor $g_{\mu\nu}$ are
\[
g_{11}=-\left(1-\frac{2M}{r} - \frac{\Lambda r^2}{3}\right)^{-1} , \quad g_{22}=- r^2,
\]
\begin{equation}\label{1}
g_{33}=-r^2\sin^2\theta, \quad
g_{44}=1-\frac{2M}{r} - \frac{\Lambda r^2}{3},
\end{equation}
where $M$ and $\Lambda>0$ are, respectively, the mass parameter and the cosmological constant (in this paper we use  the signature of the metric (--,--,--,+) and the unites $c=G=1$ are chosen).

In order to investigate the physical properties of the Schwarzschild-de Sitter background it is appropriate to consider some specific features of the gravitational field, which is determined by metric (\ref{1}), from the point of view of any observer moving relative to a source with the mass $M$. For the description of this motion one can use the local tetrad coordinates which are put in the parenthesis below. According to the general definition the expressions of the values  $E_{(k)}^{(i)}$ and $B_{(k)}^{(i)}$
are \cite{Thorne}: 
\begin{equation}\label{2}
E_{(k)}^{(i)}=R^{(i)(4)}_{}{}{}{}{}{}_{(k)(4)},
\end{equation}
\begin{equation}\label{3}
B_{(k)}^{(i)}=-\frac12 R^{(i)(4)}_{}{}{}{}{}{}_{(m)(n)}
\varepsilon^{(m)(n)}_{}{}{}{}{}{}_{(k)},
\end{equation}
where $E_{(k)}^{(i)}$ and $B_{(k)}^{(i)}$ are the gravitoelectric and gravitomagnetic components of the gravitational field respectively (in contrast to Greek indices which run 1, 2, 3, 4,  Latin indices in (\ref{2}) and (\ref{3}) run 1, 2, 3).

Let us consider relationships (\ref{2}) and (\ref{3}) in the specific case of the Schwarzschild-de Sitter metric. Without loss of generality we put that the plane of the observer motion is $\theta=\pi/2$. It is convenient to orient the first space axis (1) of the local basis along the direction which is orthogonal to the plane determined by the direction of the  observer motion and the radial direction to the mass $M$, and the second space axis (2) we orient along the direction of this motion. As a direct result of such choice of the local basis orientation we write $\vec e_{(1)}\vec e^1=0, \vec e_{(1)}\vec e^3=0, \vec e_{(2)}\vec e^2=0, \vec e_{(3)}\vec e^2=0$,
where $\vec e_{(i)}$ and $\vec e^j$ are the vectors of the local and global basis respectively. Then for the tetrad components $\lambda^\mu_{~(\nu)}$ we have 
$\lambda^1_{~(1)}=0,  \lambda^3_{~(1)}=0,  \lambda^2_{~(2)}=0,  \lambda^2_{~(3)}=0$.
Other tetrad components which are nonzero can be found from the general relationship between the metric tensor $g^{\pi\rho}$ and $\lambda^\mu_{~(\nu)}$
\begin{equation}\label{4} 
\lambda^\pi_{~(\mu)}\lambda^\rho_{~(\nu)}\eta^{(\mu)(\nu)}=g^{\pi\rho},
\end{equation}
where $\eta^{(\mu)(\nu)}$ is the Minkowski tensor. It is not difficult to verify that according to (\ref{4}) the full set of the nonzero tetrad components can be presented in the following form
$$
\lambda^2_{~(1)}=\sqrt{-g^{22}}, \quad 
\lambda^1_{~(2)}=u^1u^4\sqrt{\frac{g_{44}}{u_4u^4-1}},
$$
$$
\lambda^3_{~(2)}=u^3u^4\sqrt{\frac{g_{44}}{u_4u^4-1}}, \quad
\lambda^4_{~(2)}=\sqrt{\frac{u_4u^4-1}{g_{44}}},
$$
$$
\lambda^1_{~(3)}=u^3\sqrt{\frac{g^{11}g_{33}}{u_4u^4-1}}, \quad
\lambda^3_{~(3)}=-u^1\sqrt{\frac{g^{33}g_{11}}{u_4u^4-1}},
$$
\begin{equation}
\label{5}
\lambda^1_{~(4)}=u^1, \quad \lambda^3_{~(4)}=u^3, \quad \lambda^4_{~(4)}=u^4,
\end{equation}
where $u^1=dr/ds$, $u^3=d\varphi/ds$, $u^4=dt/ds$ are the components of the observer 4-velocity (for the motion in the plane $\theta=\pi/2$ we have $u^2=0$).

To obtain the nonzero tetrad components of Riemann's tensor $R_{(\pi)(\rho)(\sigma)(\tau)}$ we use 
the general relationship
\begin{equation}
\label{6}
R_{(\pi)(\rho)(\sigma)(\tau)}=
\lambda^\alpha_{~(\pi)}\lambda^\beta_{~(\rho)}
\lambda^\gamma_{~(\sigma)}\lambda^\delta_{~(\tau)}
R_{\alpha\beta\gamma\delta},
\end{equation}
where $R_{\alpha\beta\gamma\delta}$ are the components of the Riemann tensor in the global coordinates. The components $R_{\alpha\beta\gamma\delta}$ for metric (\ref{1}) can be calculated in the direct manner and for the sake of brevity we do not write here the corresponding explicit expressions. Instead of taking into account relationships (\ref{5}) and (\ref{6}) we write the final nonzero expressions for $E_{(k)}^{(i)}$ and $B_{(k)}^{(i)}$ according to (\ref{2}) and (\ref{3})
$$
E^{(1)}_{(1)}=\frac{M}{r^3} - \frac{\Lambda}{3} +\frac{3M}{r^3} u_\perp^2,
$$
$$
E^{(2)}_{(2)}=-\frac{2M}{r^3} - \frac{\Lambda}{3} +\frac{3M}{r^3}\frac{u_\perp^2}{u_4u^4-1},
$$
$$
E^{(2)}_{(3)}=E^{(3)}_{(2)}=-\frac{3M}{r^3}\frac{u_\parallel u_\perp u^4}
{u_4u^4-1},
$$
\begin{equation}\label{7}
 E^{(3)}_{(3)}=\frac{M}{r^3} - \frac{\Lambda}{3} -
\frac{3M}{r^3}\frac{u_\perp^2 u_4 u^4}{u_4u^4-1},
\end{equation}
$$
B^{(1)}_{(2)}=B^{(2)}_{(1)}=
\frac{3M u_\parallel u_\perp}
{r^3\sqrt{u_4u^4-1}}\left(1-\frac{2M}{r} - \frac{\Lambda r^2}{3}\right)^{-1/2},
$$
\begin{equation}\label{8}
B^{(1)}_{(3)}=B^{(3)}_{(1)}=
\frac{3M u_\perp^2 u^4}
{r^3\sqrt{u_4u^4-1}}\left(1-\frac{2M}{r} - \frac{\Lambda r^2}{3}\right)^{1/2},
\end{equation}
where $u_{\parallel} = dr/ds$ and 
$u_{\perp} = r d\varphi/ds$ are the radial and tangential components of the observer 4-velocity, and by the general expression $u_\mu u^\mu=1$ we have 
\begin{equation}\label{9}
u_4u^4-1=u_\perp^2
+\left(1-\frac{2M}{r} - \frac{\Lambda r^2}{3}\right)^{-1}u_\parallel^2.
\end{equation}
Relationships (\ref{7}) and (\ref{8}) hold true for any arbitrary velocity of the observer if the condition 
\begin{equation}\label{10}
1-\frac{2M}{r} - \frac{\Lambda r^2}{3}>0,
\end{equation}
i.e. $g_{44}>0$, is satisfied.

Similarly to the case of Schwarzschild's background which was considered in
\cite{Mash-92}, expressions (\ref{7}) and (\ref{8}) show that the special direction exists in space along which the components $E_{(k)}^{(i)}$ and $B_{(k)}^{(i)}$ remain finite because in the  case with
$u_\perp =0$ all components $E_{(k)}^{(i)}$ and $B_{(k)}^{(i)}$ are finite for any $u_\parallel$. Moreover, then according to
(\ref{8}) all components $B_{(k)}^{(i)}$ are equal to 0 and this property is similar to the corresponding one in electrodynamics when the usual magnetic field of the moving electric charge is estimated.

In general, when $u_\perp \ne 0$, the components $E_{(k)}^{(i)}$ and $B_{(k)}^{(i)}$ in (\ref{7}) and (\ref{8}) significantly depend on the observer motion. Note that the corresponding dependence of the gravitoelectric components in Schwarzschild's field as well as the gravitomagnetic components caused by a rotating mass (in the post-Newtonian approximation) were considered in \cite{Mash-92} in the context of the tidal accelerations estimation.

It is convenient to consider expressions (\ref{7}) and (\ref{8}) using the Lorentz factor $\gamma$ as estimated by an observer which is at rest relative to the source of the gravitational field. The value of $\gamma$ which corresponds to the  $u_\parallel$ and $u_\perp$ is given by the expression
\begin{equation}\label{cor1}
\gamma=\frac{1}{\sqrt{1-v^2}},
\end{equation}
where $v^2$ is the second power of the particle's 3-velocity relative to the observer. In the case of the diagonal metric, according to the general expression for the 3-velocity components
$v^i$ we have \cite{Landau}
\begin{equation}\label{cor2}
v^i=\frac{dx^i}{\sqrt{g_{44}}dt}.
\end{equation}
Then for $v^2$ we write
\begin{equation}\label{cor3}
v^2=v_i v^i = \gamma_{ik}v^i v^k,
\end{equation}
where $\gamma_{ik}$ is the $3$-space metric tensor, with the following relationship between $\gamma_{ik}$ and $g_{\mu\nu}$ for the diagonal metric:
$\gamma_{ik}$=-$g_{ik}$. It follows from (\ref{cor1})--(\ref{cor3})
with $u_\mu u^\mu=1$ that
\begin{equation}\label{cor4}
\gamma = \sqrt{u_4 u^4}.
\end{equation}

For our further purposes we write the values of the components $B_{(k)}^{(i)}$  using $\gamma$- factor explicitly. Then by (\ref{8}) and (\ref{cor4}) we have
\begin{equation}\label{cor5}
B^{(1)}_{(2)}=B^{(2)}_{(1)}=
\frac{3M}
{r^3}\frac{u_\parallel u_\perp}{\sqrt{\gamma^2 -1}}\left(1-\frac{2M}{r} - \frac{\Lambda r^2}{3}\right)^{-1/2},
\end{equation}
\begin{equation}\label{cor6}
B^{(1)}_{(3)}=B^{(3)}_{(1)}=
\frac{3M}
{r^3}\frac{u_\perp^2 \gamma}{\sqrt{\gamma^2 -1}}.
\end{equation}
Let us compare the values from (\ref{cor5}) and (\ref{cor6}) at low and high velocities. When the velocity is low with $u_\parallel = \delta_1$, $u_\perp = \delta_2$, $ |\delta_1|\ll 1$, $ |\delta_2|\ll 1$, and 
$\gamma^2-1 = \Delta^2\ll 1$, where by (\ref{9})
\begin{equation}\label{cor7}
\Delta^2=\left(1-\frac{2M}{r} - \frac{\Lambda r^2}{3}\right)^{-1}\delta_1^2 + \delta_2^2,
\end{equation}
it follows from (\ref{cor5}) and (\ref{cor6}) that
\begin{equation}\label{cor8}
B^{(1)}_{(2)}=B^{(2)}_{(1)}\approx
\frac{3M}
{r^3}\frac{\delta_1 \delta_2}{\Delta}\left(1-\frac{2M}{r} - \frac{\Lambda r^2}{3}\right)^{-1/2},
\end{equation}
\begin{equation}\label{cor9}
B^{(1)}_{(3)}=B^{(3)}_{(1)}\approx
\frac{3M}
{r^3}\frac{\delta_2^2}{\Delta}.
\end{equation}
That is, at low velocity the common term $3M/r^3$ in the expressins for the gravitomagnetic components  (\ref{cor8}) and (\ref{cor9}) is multiplied by corresponding small factors. Whereas in the highly relativistic region, when 
$\gamma^2\gg 1$ and both $u_\parallel^2$ and $u_\perp^2$ have order
$\gamma^2$, it follows from  (\ref{cor5}) and (\ref{cor6}) that
\begin{equation}\label{cor10}
B^{(1)}_{(2)}=B^{(2)}_{(1)}\sim
\frac{3M}
{r^3}\left(1-\frac{2M}{r} - \frac{\Lambda r^2}{3}\right)^{-1/2}\gamma,
\end{equation}
\begin{equation}\label{cor11}
B^{(1)}_{(3)}=B^{(3)}_{(1)}\sim
\frac{3M}
{r^3}\gamma^2.
\end{equation}
When only  $u_\perp^2\gg 1$, with $u_\parallel^2\ll u_\perp^2$, the values from (\ref{cor10}) are proportional to $u_\parallel$, and the values from (\ref{cor11}) are proportional to $\gamma^2$. In the case, when
$u_\parallel^2\gg 1$ and $u_\perp^2\ll u_\parallel^2$,  the values from (\ref{cor10}) and (\ref{cor11}) are proportional to $u_\perp$ and $u_\perp^2$, respectively.

Note that in contrast to the dependence on $\gamma$ of the magnetic field of a moving electric charge in electrodynamics, where  components of this field are proportional to $\gamma$,  some above considered gravitomagnetic components can be proportional to  $\gamma$ and others to $\gamma^2$.

\section{Spin-gravity-$\Lambda$ coupling by estimation of an observer comoving with the spinning particle}

\subsection{Some general relationships connected with the MP equations}

We take into account the MP equations in the form  \cite{Mathis, Papa}
\begin{equation}\label{1a}
\frac D {ds} \left(mu^\lambda + u_\mu\frac {DS^{\lambda\mu}}
{ds}\right)= -\frac {1} {2} u^\pi S^{\rho\sigma}
R^{\lambda}_{~\pi\rho\sigma},
\end{equation}
\begin{equation}\label{2a}
\frac {DS^{\mu\nu}} {ds} + u^\mu u_\sigma \frac {DS^{\nu\sigma}}
{ds} - u^\nu u_\sigma \frac {DS^{\mu\sigma}} {ds} = 0,
\end{equation}
where $u^\lambda\equiv dx^\lambda/ds$ is the particle's 4-velocity,
$S^{\mu\nu}$ is the tensor of spin, $m$ and $D/ds$ are the mass and the covariant derivative along $u^\lambda$, respectively. 

As usual, these equations are considered with some supplementary condition, and more often than not either the Mathisson-Pirani condition \cite{Mathis, Pirani}
\begin{equation}\label{3a}
S^{\lambda\nu} u_\nu = 0
\end{equation}
or the Tulczyjew-Dixon one \cite{Tul, Dixon}
\begin{equation}\label{4a}
S^{\lambda\nu} P_\nu = 0
\end{equation}
are used, where
\begin{equation}\label{5a}
P^\nu = mu^\nu + u_\lambda\frac{DS^{\nu\lambda}}{ds}
\end{equation}
is the particle 4-momentum. In both  (\ref{3a}) and (\ref{4a}), the constant of motion of the MP equations is
\begin{equation}\label{6a}
 S_0^2=\frac12 S_{\mu\nu}S^{\mu\nu},
\end{equation}
where $|S_0|$ is the absolute value of spin. The value $m$ in the left-hand side of Eq. (\ref{1a}) is the constant of motion under condition (\ref{3a}).
The value $P_\mu P^\mu$ is the constant of motion at condition (\ref{4a}). 

In recent papers  \cite{Ohashi, Kyr, Semerak} instead of Eqs. (\ref{1a}) and (\ref{2a}) the set of equations 
\begin{equation}\label{1aa}
m\frac {Du^\lambda} {ds} = -\frac {1} {2} u^\pi S^{\rho\sigma}
R^{\lambda}_{~\pi\rho\sigma},
\end{equation}
\begin{equation}\label{2aa}
\frac {DS^{\mu\nu}} {ds} = 0,
\end{equation}
\begin{equation}\label{3aa}
\frac {Dm} {ds} = 0
\end{equation}
is studied and it means that the new Ohashi-Kyrian-Semer{\'a}k supplementary condition is used. This approach is related to the possibility to exclude the hidden momentum which is determined by the
term
$$
u_\mu\frac {DS^{\lambda\mu}}
{ds}
$$
in the MP equations (\ref{1a}) and (\ref{2a}).

Other aspects of the supplementary conditions for the MP equations are elucidated, for example, in \cite{Costa, Mik}. As it is pointed out in \cite{Wald}, the physical condition for a spinning test particle 
\begin{equation}\label{6b}
 \varepsilon \equiv \frac{|S_0|}{mr}\ll 1
\end{equation}
must be taken into account.

In many papers where condition (\ref{3a}) is used  the 4-vector of spin $s_\lambda$ is taken into account in the MP equations, instead of the 4-tensor $S^{\mu\nu}$, where by definition
\begin{equation}\label{7a}
s_\lambda=\frac{1}{2} \sqrt{-g} \varepsilon_{\lambda\mu\nu\sigma} u^\mu S^{\nu\sigma},
\end{equation}
$g$ is the determinant of $g_{\mu\nu}$ and $\varepsilon_{\lambda\mu\nu\sigma}$ is the Levi-Civita symbol. Under condition (\ref{3a}) it follows from (\ref{2a}) that 
\begin{equation}\label{7b}
\frac{Ds^\lambda}{ds}=s_\mu\frac{Du^\mu}{ds}u^\lambda,
\end{equation}
and it means that the 4-vector of spin is Fermi transported.
The relationship $s_\lambda s^\lambda = S_0^2$ holds.
In addition, in practical calculations it is convenient to represent the MP equations through the spin 3-vector $S_i$ which is defined by
\begin{equation}\label{8a}
S_i=\frac{1}{2} \sqrt{-g} \varepsilon_{ikl}  S^{kl}.
\end{equation}
The following relationship holds:
\begin{equation}\label{9a}
S_i= u_i s_4 - u_4 s_i.
\end{equation}
At condition (\ref{3a}) the three independent equations of set (\ref{2a}) can be written as
$$
u_4 \dot S_i - \dot u_4 S_i +  2(\dot u_{[4} u_{i]} -
u^\pi u_\rho \Gamma^\rho_{\pi[4} u_{i]})S_k u^k
$$
\begin{equation}\label{9b}
+ 2S_n \Gamma^n _{\pi [4} u_{i]} u^\pi =0,
\end{equation}
where a dot denotes the usual differentiation with respect to the proper time $s$, and square brackets denote antisymmetrization of indices. 

In general, the solutions of Eqs. (\ref{1a}) and (\ref{2a}) under conditions
(\ref{3a}) and (\ref{4a}) are different. However, in the post-Newtonian approximation these solutions coincide with high accuracy \cite{Bark}, just as in other cases when the spin effects lead to small corrections to the corresponding geodesic expressions. Therefore, instead of rigorous MP Eqs.
 (\ref{1a}) their linear spin approximation
\begin{equation}\label{10a}
m\frac{D}{ds} u^\lambda = -\frac{1}{2} u^\pi S^{\rho\sigma}
R^{\lambda}_{~\pi\rho\sigma}
\end{equation}
is often considered. In this approximation condition (\ref{4a}) matches with (\ref{3a}).

An important physical consequence follows from Eqs. (\ref{10a}), after their representation in  terms of the local (tetrad) values, which describe the situation when an observer is comoving with the spinning particle. Then one can obtain from Eqs. (\ref{10a}) the relationships \cite{Pl98, Pl15}
\begin{equation}\label{plyatsko:20}
\gamma_{(i)(4)(4)} = -\frac{s_{(1)}}{m} R_{(i)(4)(2)(3)},
\end{equation}
where $\gamma_{(k)(1)(4)}$ are the Ricci coefficients of rotation, and here
the first local vector (1) is chosen to be oriented along the particle's spin (it means that $s_{(2)}=0$, $s_{(3)}=0$, and by the property of the spin 4-vector for a comoving observer $s_{(4)} = 0$ \cite{Mash-71}).

It is known that value $\gamma_{(i)(4)(4)}$ is the dynamical characteristic of the reference frame, namely, its acceleration. That is, according to (\ref{plyatsko:20}) we have
\begin{equation}\label{plyatsko:21}
a_{(i)} = -\frac{s_{(1)}}{m} R_{(i)(4)(2)(3)},
\end{equation}
where $a_{(i)}$ is the 3-acceleration with which the spinning particle deviates from free geodesic fall as measured by the comoving observer. Taking (\ref{3}) into account it is easy to see that the right-hand side of (\ref{plyatsko:21}) contains the gravitomagnetic components.

For comparison, we present the relationship which follows from the rigorous MP equations (\ref{1a}) under condition (\ref{3a}). Taking into account the condition of the Fermi transport for the local orthogonal vectors 
$\gamma_{(i)(k)(4)}=0$ we write this relationship as \cite{Pl15}
\begin{equation}\label{plyatsko:21a}
a_{(1)} = -\frac{s_{(1)}}{m} R_{(1)(4)(2)(3)},
\end{equation}
\begin{equation}\label{plyatsko:21b}
a_{(2)} = -\frac{s_{(1)}}{m}(R_{(2)(4)(2)(3)} - \dot a_{(3)}),
\end{equation}
\begin{equation}\label{plyatsko:21c}
a_{(3)} = -\frac{s_{(1)}}{m}(R_{(3)(4)(2)(3)} - \dot a_{(2)}),
\end{equation}
where a dot denotes the usual derivative with respect to $s$. In a particular case of $a_{(2)}=const$ and $a_{(3)}=const$ Eqs. (\ref{plyatsko:21a})--(\ref{plyatsko:21c}) coincide with Eq. (\ref{plyatsko:21}). Some corresponding special examples  will be considered below.

\subsection{The acceleration $a_{(i)}$ in the Schwarzschild-de Sitter background}

Let us consider (\ref{plyatsko:21}) in the specific case of spinning particle motion in the Schwarzschild-de Sitter background, when the particle's spin is orthogonal to the plane determined by the direction of the particle motion and the radial direction. We choose the same orientation of the space local axes as in the previous section. Then by (\ref{plyatsko:21}) we write
\begin{equation}\label{plyatsko:22}
a_{(i)} = -\frac{s_{(1)}}{m} B^{(1)}_{(i)}.
\end{equation}
Because $B^{(1)}_{(i)}=0$ (all nonzero components  $B^{(i)}_{(k)}$ are
written in (\ref{8})), we have $a_{(1)}=0$, i.e the acceleration is absent in the direction of the spin orientation. According to (\ref{8}) and (\ref{plyatsko:22}) the nonzero components $a_{(2)}$ and $a_{(3)}$ are
\begin{equation}\label{plyatsko:23}
a_{(2)}= -\frac{s_{(1)}}{m}
\frac{3M u_\parallel u_\perp}
{r^3\sqrt{u_4u^4-1}}\left(1-\frac{2M}{r} - \frac{\Lambda r^2}{3}\right)^{-1/2},
\end{equation}
\begin{equation}\label{plyatsko:24}
a_{(3)}= -\frac{s_{(1)}}{m}
\frac{3M u_\perp^2 u^4}
{r^3\sqrt{u_4u^4-1}}\left(1-\frac{2M}{r} - \frac{\Lambda r^2}{3}\right)^{1/2}.
\end{equation}
For the absolute value of the acceleration $|\vec a|$,
where
$$
|\vec a| = \sqrt{a_{(1)}^2 + a_{(2)}^2 + a_{(3)}^2},
$$
 according to (\ref{plyatsko:23}), (\ref{plyatsko:24}), and (\ref{9}) we get
\begin{equation}\label{plyatsko:25}
|\vec a|= \frac{3M}{r^2}
\frac{|S_0|}{mr} |u_\perp|\sqrt{1+u_\perp^2}.
\end{equation}
In (\ref{plyatsko:25}) we take into account that $|s_{(1)}|=|S_0|$ (it follows from the general relationship $s_{(\mu)}s^{(\mu)}=s_{\mu} s^{\mu} = S_0^2)$. One can check that the vector $\vec a$ is oriented along the radial direction, because the scalar product of $\vec a$ and the tangentially directed vector $\vec e_3$ is equal to 0. Indeed, 
$$
\vec a = a_{(2)}\vec e^{(2)} + a_{(3)}\vec e^{(3)},
$$
then 
$$
\vec a\vec e_3=(a_{(2)}\vec e^{(2)} + a_{(3)}\vec e^{(3)})\vec e_3=
a_{(2)}\lambda_3^{~(2)} + a_{(3)}\lambda_3^{~(3)}
$$
\begin{equation}\label{plyatsko:26}
= -g_{33}(a_{(2)}\lambda^3_{~(2)} + a_{(2)}\lambda^3_{~(2)}).
\end{equation}
Using expressions (\ref{5}), (\ref{plyatsko:23}), and (\ref{plyatsko:24}) we obtain that the value in the last brackets of (\ref{plyatsko:26}) is equal to 0.

We also emphasize that the value of acceleration (\ref{plyatsko:25}) does not depend on the radial component of the particle velocity $u_{\parallel}$ and depends significantly on its tangential velocity $u_{\perp}$. That is, expression (\ref{plyatsko:25}) shows the essential difference between the two cases, when 1. $u_{\perp}\ll1$ (the weak relativistic motion), and 2.  $u_{\perp}\gg 1$ (the highly relativistic motion). In the first case after (\ref{plyatsko:25}) we write
\begin{equation}\label{plyatsko:27}
|\vec a|=\frac{3M}{r^2}\varepsilon \delta,
\end{equation}
where the small value $\varepsilon$ is determined in (\ref{6b}), and $\delta\equiv  |u_\perp|\ll 1$. Note that $M/r^2$ is equal to the Newtonian acceleration of  free fall which is caused by a body with the mass $M$. According to  (\ref{plyatsko:27}) the acceleration of a spinning particle is much less than the Newtonian value $M/r^2$, whereas in the second case by  (\ref{plyatsko:25}) we have
\begin{equation}\label{plyatsko:28}
|\vec a|= \frac{3M}{r^2}\varepsilon \gamma^2,
\end{equation}
where $\gamma$ is the Lorentz factor calculated by the tangential velocity
$u_\perp$. Expression (\ref{plyatsko:28}) shows that for any small value $\varepsilon$ one can choose such the high values $\gamma$ which would lead to $|\vec a|\gg M/r^2$.

Note that expression (\ref{plyatsko:25}) is common for Schwarzschild's and the Schwarzschild-de Sitter background and the role of the $\Lambda$ can be revealed by consideration of the dependence of $u_\perp$ on $\Lambda$ for specific spinning particle motions. Naturally, it is important to study the cases of motions which can be described in some clear analytic expressions, if it is possible. For this purpose in the next section we consider the circular orbits of a spinning particle, when $u_\parallel=0$ and $u_\perp \ne 0$. In this context we note that on the circular orbits with $u_\perp = const$ expression (\ref{plyatsko:20}) is rigorous, in contrast to the noncircular orbits when (\ref{plyatsko:20}) is within the linear spin approximation. In the following we will consider rigorous MP equations 
(\ref{1a}) and (\ref{2a}) under condition (\ref{3a}).

\section{Highly relativistic circular orbits of a spinning particle in the Schwarzschild-de Sitter spacetime}

First, note that by the geodesic equations in metric (\ref{1}) circular orbits of a spinless particle the Schwarzschild-de Sitter spacetime are allowed when its orbital (tangential) 4-velocity $u_{\perp}$ satisfies the relationship
\begin{equation}\label{n}
u_{\perp}^2=\left(\frac{M}{r}-\frac{\Lambda}{3}r^2\right)\left(1-\frac{3M}{r}\right)^{-1}.
\end{equation}
It follows from (\ref{n}) that under the condition
\begin{equation}\label{n+1}
\frac{M}{r}>\frac{\Lambda}{3}r^2
\end{equation}
circular orbits exist  for $r>3M$, similarly as for Schwarzschild's spacetime, and the orbit with $r=3M$ is the so called orbit of a photon. The circular orbits are highly relativistic only at  $r=3M(1+\delta)$, where $0<\delta \ll 1$, i.e. in a small neighborhood of the value $3M$.
Note that the right-hand side of (\ref{n}) is also positive when
\begin{equation}\label{n+1a}
\frac{M}{r}<\frac{\Lambda}{3}r^2, \quad r<3M.
\end{equation}
However, it follows from (\ref{n+1a}) that then $g_{44}<0$, i.e. then the meaning of the coordinate $t$ is changed.

Now we take into account the MP equations (\ref{1a}) and (\ref{2a}) with 
supplementary condition (\ref{3a}) because, in general, for a correct description of highly relativistic motions of a spinning particle this condition is more appropriate than condition (\ref{4a}) \cite{Pl12, Pl15, Pl16b}. (Some points pertaining to the issue of choosing different supplementary conditions are discussed below in subsection A).
It is easy to check by direct calculations that in the case when spin is orthogonal to the plane $\theta=\pi/2$ in metric (\ref{1}), i.e. when
 $S_1\equiv S_r=0$,
$S_3\equiv S_\varphi=0$, and only $S_2\equiv S_\theta$ is nonzero,
Eqs. (\ref{1a})-(\ref{3a}) in this metric have solutions which describe the circular orbits of a spinning particle in the plane $\theta=\pi/2$. According to these solutions, there is the connection between the nonzero components of the particle 4-velocity 
$u^3$, $u^4$ and the component of the 3-vector of spin $S_2$: 
$$
- r\left[u^3 u^3 - \left(\frac{M}{r^3} - \frac{\Lambda}{3}\right)u^4 u^4\right]
$$
$$
\times \left[1- \left(1-\frac{3M}{r}\right)\left(1-\frac{2M}{r} - \frac{\Lambda r^2}{3}\right)^{-1} u^3 \frac{S_2}{mr}\right]
$$
\begin{equation}\label{n+2}
=\frac{3M}{r^3}u^3 \left(1-\frac{2M}{r} - \frac{\Lambda r^2}{3}\right)^{-1} \frac{S_2}{mr}
\end{equation}
(relationship (\ref{n+2}) follows directly from the first equation from (\ref{1a}), with $\lambda = 1$).
In addition to (\ref{n+2}), it is necessary to take into account the relationship for $u^3$ and $u^4$
\begin{equation}\label{n+3}
- r^2 u^3 u^3 + \left(1-\frac{2M}{r} - \frac{\Lambda r^2}{3}\right) u^4 u^4 = 1,
\end{equation}
which is a direct consequence of the general relationship $u_\mu u^\mu=1$. From  (\ref{n+2}) and  (\ref{n+3}) we obtain the third order algebraic equation for $u_\perp$
$$
u_\perp^3 \left(1-\frac{3M}{r}\right)^2 \frac{S_2}{mr^2} -
u_\perp^2 \left(1-\frac{3M}{r}\right)\left(1-\frac{2M}{r} - \frac{\Lambda r^2}{3}\right)
$$
$$
+ u_\perp \frac{S_2}{mr^2}\left[\frac{M}{r}\left(2-\frac{3M}{r}\right) +
\frac{\Lambda r^2}{3}\left(1-\frac{6M}{r}\right)\right]
$$
\begin{equation}\label{n+4}
+\left(\frac{M}{r} - \frac{\Lambda r^2}{3}\right) \left(1-\frac{2M}{r} - \frac{\Lambda r^2}{3}\right) = 0,
\end{equation}
which determines the dependence of the spinning particle velocity on $r$,
$M$, $\Lambda$, and $S_2$. Without any loss in generality,  we can put $S_2>0$. By the way, the set of equations for the spin 3-vector (\ref{9b}) becomes much simpler when spin is orthogonal to the trajectory of motion ($S_i u^i=0$):
\begin{equation}\label{9c}
u_4 \dot S_i - \dot u_4 S_i + 2S_n \Gamma^n _{\pi [4} u_{i]} u^\pi =0.
\end{equation}
In our case of  particle motions in the equatorial plane of the Schwarzschild-de Sitter metric with $S_1=0$ and $S_3=0$ the single nontrivial equation from (\ref{9c}) with $i=2$ can be easily integrated. As a result, we have 
\begin{equation}\label{9d}
S_2 = ru_4 S_0.
\end{equation}

\subsection{The case $r=3M$}

In the particular case, when $r=3M$, equation (\ref{n+4}) becomes linear for $u_\perp$ and we have 
\begin{equation}\label{n+5}
u_\perp = - \frac{3mM^2}{S_2}(1-9\Lambda M^2).
\end{equation}
Without any loss in generality,  we can put $S_2>0$. Then after (\ref{9d})  using (\ref{n+3}) we rewrite (\ref{n+5}) in the form
\begin{equation}\label{n+6}
u_\perp \approx - \left(\frac{1}{3} - 3\Lambda M^2\right)^{1/4}\frac{1}{\sqrt{\varepsilon}}\left(1- \frac{\varepsilon}{4}\left(\frac{1}{3} - 3\Lambda M^2\right)^{-1/2} \right)
\end{equation}
where $\varepsilon$ is determined in (\ref{6b}). It follows from (\ref{6b}) and  (\ref{n+6})   that 
\begin{equation}\label{n+7}
u_\perp^2 \gg 1
\end{equation}
and it means that for the motion on the circular orbit with $r=3M$ the spinning particle must posses a highly relativistic velocity.

There are two points here worth drawing our attention to. The first one is that the term in the left-hand side of (\ref{n+2})
\begin{equation}\label{n+8}
\left(1-\frac{3M}{r}\right)\left(1-\frac{2M}{r} - \frac{\Lambda r^2}{3}\right)^{-1} u^3 \frac{S_2}{mr}
\end{equation}
is determined just by the term 
\begin{equation}\label{n+8a}
u_\mu\frac {DS^{\lambda\mu}}
{ds}
\end{equation}
from the left-hand side of (\ref{1a}) and in the right-hand side of (\ref{5a}). The second one is that for the  considered circular orbit with $r=3M$ both term (\ref{n+8}) and (\ref{n+8a}) are equal to 0. They allow us to make a conclusion that partial solution  (\ref{n+5}) of the the MP equations in the Schwarzschild-de Sitter background which describes the highly relativistic circular orbit with $r=3M$ is common under the Mathisson-Pirani, Tulczyjew-Dixon, and
Ohashi-Kyrian-Semer{\'a}k conditions. From this point this solution is unique (nevertheless, it cannot be excluded that the MP equations may have some other solutions that describe  noncircular highly relativistic orbits and that are the same or close under different supplementary conditions). With this in mind, it should also be noted that in order to describe motion of a spinning particle correctly any partial solution of the MP equations which is obtained under some supplementary condition other than the  Mathisson-Pirani one has to be the same or very close to the corresponding nonhelical solution of the MP equations under the Mathisson-Pirani condition (this point is elucidated in \cite{Pl12, Pl15}).

It is appropriate to compare solution (\ref{n+5}) with the well-known circular solutions of the exact  MP equations under condition (\ref{3a}) in the Minkowski spacetime. From Eq. (\ref{n+5}) in the case of $M=0$ and $\Lambda=0$ we have 
\begin{equation}\label{n+5a}
u_\perp = \frac{mr^2}{S_2}.
\end{equation}
(Circular solution (\ref{n+5a}) represents a partial case of the general helical solutions which have Eqs. (\ref{1a})-(\ref{3a}) in the Minkowski spacetime.) As it can easily be seen, expressions (\ref{n+5}) and (\ref{n+5a}) have different signs for the fixed sign of $S_2$. It means that the directions of the rotation according to (\ref{n+5}) and (\ref{n+5a}) are opposite, and that the circular orbit (\ref{n+5}) cannot be interpreted as the one which is analogous to  the orbits with  (\ref{n+5a}) (Eq. (\ref{n+5a}) does not follow from Eq. (\ref{n+5}) when $M$ tends to 0 for $\Lambda =0$).

\subsection{The case $r<3M$, $\frac{M}{r} - \frac{\Lambda r^2}{3} > 0$}  

In this case the cubic Eq. (\ref{n+4}) has the single real root which is
$$
u_\perp \approx - \left(\frac{3M}{r} - 1\right)^{-1/2} \left(1-\frac{2M}{r} - \frac{\Lambda r^2}{3}\right)^{1/4}
$$
\begin{equation}\label{n+9}
\times \frac{1}{\sqrt{\varepsilon}}(1+k\varepsilon),
\end{equation}
where
\begin{equation}\label{n+9a}
k=\frac{c}{d} - \frac{1}{4}\left(1- \frac{3M}{r}\right)\left(1-\frac{2M}{r} - \frac{\Lambda r^2}{3}\right)^{-1/2},
\end{equation}
$$
c=-\left(\frac{M}{r} - \frac{\Lambda r^2}{3}\right)
\left[\frac{M}{r}\left(5-12 \frac{M}{r}\right) - \frac{\Lambda r^2}{3}
\left(2-3 \frac{M}{r}\right)\right]^3
$$
$$
+ \left[\frac{M}{r}\left(2-3 \frac{M}{r}\right) + \frac{\Lambda r^2}{3}
\left(1-6 \frac{M}{r}\right)\right]^2
$$
$$
\times \left[\frac{M}{r}\left(4-3 \frac{M}{r}- 18\frac{M^2}{r^2}\right)\right.
 $$
$$
+ \left.\frac{\Lambda r^2}{3}
\left(2-33 \frac{M}{r}+ 72\frac{M^2}{r^2}\right) + \frac{\Lambda^2 r^4}{3}\left(2-3 \frac{M}{r}\right) \right],
$$
$$
d=54\left|1-\frac{3M}{r}\right|^3 \left(\frac{M}{r} - \frac{\Lambda r^2}{3}\right)^3 \left(1-\frac{2M}{r} - \frac{\Lambda r^2}{3}\right)^{1/2},
$$
and these expressions are written for the case when
$$
\varepsilon \left|1-\frac{3M}{r}\right|^{-3} \ll 1,
$$
i.e. when $r$ is not very close to $3M$.
Similarly to the case with $r=3M$, here the value $u_{\perp}$ is proportional to $1/\sqrt{\varepsilon}$ and $u_{\perp}<0$ for the chosen positive sign of $S_2$. Note that the case of  highly relativistic circular orbits with $r<3M$ in Schwarzschild's background was under investigation in \cite{Pl05, Pl12}. It was shown that these orbits exist due to the strong coupling of spin with the gravitational field and the force of this coupling acts on a particle as a repulsive one \cite{Pl12}. (It is known that a spinless particle, which starts in the tangential direction relative to Schwarzschild's mass with any velocity from the position $r<3M$, falls on the horizon surface). The specific feature of the Schwarzschild-de Sitter background is the property that the constant $\Lambda>0$ describes the cosmological repulsion. According to (\ref{n+9}), the presence of $\Lambda>0$ in the right-hand side of (\ref{n+9}) leads to less absolute value of the tangential velocity (and the corresponding $\gamma$-factor), which is necessary for the realization of the circular orbits with some fixed $r<3M$, than it is necessary in Schwarzschild's background.

As it is noted above the case 
$$
r<3M, \quad \frac{M}{r} - \frac{\Lambda r^2}{3} < 0
$$
means that $g_{44}<0$.

\subsection{The case $r>3M$, $\frac{M}{r} - \frac{\Lambda r^2}{3} > 0$}

Under these conditions Eq. (\ref{n+4}) has three real roots for $u_\perp$ which can be written as
$$
u_\perp \approx \left(1- \frac{3M}{r}\right)^{-1/2} \left(1-\frac{2M}{r} - \frac{\Lambda r^2}{3}\right)^{1/4}
$$
\begin{equation}\label{n+10}
\times \frac{1}{\sqrt{\varepsilon}}(1+k\varepsilon),
\end{equation}
\begin{equation}\label{n+11}
u_{\perp 1,2} \approx \pm \left(\frac{M}{r}-\frac{\Lambda}{3}r^2\right)^{1/2}\left(1-\frac{3M}{r}\right)^{-1/2}(1 \pm \varkappa\varepsilon)),
\end{equation}
where the expression for $k$ is the same as in (\ref{n+9a})
and the expression for $\varkappa$ is
$$
\varkappa=\frac{3M}{2r}\left(1-\frac{3M}{r}\right)^{-1}
$$
\begin{equation}\label{n+11a}
\times\left(1-\frac{2M}{r} - \frac{\Lambda r^2}{3}\right)
 \left(\frac{M}{r}-\frac{\Lambda}{3}r^2\right)^{-1/2}.
\end{equation}
Concerning Eq. (\ref{n+10}) we note that in contrast to expression (\ref{n+9}) the sign of $u_\perp$ in (\ref{n+10}) is positive for the positive sign of $S_2$. In the particular case of Schwarzschild's background ($\Lambda=0$), the analogous to  (\ref{n+10})  expression was considered in \cite{Pl12} and it was pointed out that the corresponding highly relativistic circular orbits at $r>3M$ are caused by the strong attractive action of the spin-gravity coupling. If $\Lambda \ne 0$, the second bracket in the right-hand side of Eq. (\ref{n+10}) shows the influence of $\Lambda$ on the necessary value of $u_\perp$ for the realization of these orbits.

Relationship (\ref{n+11}) describes a simple situation of  circular motions of a spinning particle which are close to the corresponding geodesic motions (in this context see Eq. (\ref{n})), i.e. when the influence of the particle's spin on its orbital velocity $u_\perp$ is weak.

\subsection{The case $r>3M$, $\frac{M}{r} - \frac{\Lambda r^2}{3} < 0$}

This case does not have an analogy in Schwarzschild's background and then Eq. (\ref{n+4}) has the single real positive root for $u_\perp$:
$$
u_\perp \approx \left(1- \frac{3M}{r}\right)^{-1/2} \left(1-\frac{2M}{r} - \frac{\Lambda r^2}{3}\right)^{1/4}
$$
\begin{equation}\label{n+12}
\times \frac{1}{\sqrt{\varepsilon}}(1+ k\varepsilon)),
\end{equation}
where the expression for $k$ is the same as in (\ref{n+9a}).
The corresponding circular orbits of a spinning particle exist because of the attractive action of the spin-gravity coupling. For comparison note that according to (\ref{n}) any circular orbits of a spinless particle are possible at the conditions 
$$
r>3M, \quad \frac{M}{r} < \frac{\Lambda r^2}{3}.
$$

It is important to emphasize that from (\ref{n+6}), (\ref{n+9}), (\ref{n+10}), and (\ref{n+12}) we can see that for corresponding circular orbits of a spinning particle the absolute values of $u_\perp$ are proportional to $1/\sqrt{\varepsilon}$, and by (\ref{6b}) the relationship $u_\perp^2\gg 1$ takes place. Then in this context we point out that  according to  (\ref{plyatsko:28}) the value of the 
acceleration $|\vec a|$ is of order $M/r^2$.

Now, there are several remarks regarding the possibility of interpreting the highly relativistic circular solutions which are described by  (\ref{n+6}), (\ref{n+9}), (\ref{n+10}), and (\ref{n+12}) as corresponding to the partial case of helical solutions. First, let us take into account Eqs. (\ref{plyatsko:21b}) and (\ref{plyatsko:21c}). The general solution of these equations is
$$
a_{(2)}(s) = \frac{S_0}{M} \int\limits_0^s \left(- \dot R_{(3)(4)(2)(3)} -
\frac{M}{S_0}R_{(2)(4)(2)(3)}\right)
$$
$$
\times \sin\frac{M}{S_0}(s-x)\,dx + a_{(2)}(0) \cos\frac{M}{S_0}
$$
\begin{equation}\label{n+13}
 - \left(a_{(3)}(0) +\frac{S_0}{M}
R_{(3)(4)(2)(3)}(0)\right)\sin\frac{M}{S_0}s,
\end{equation}
$$
a_{(3)}(s) = \frac{S_0}{M} \int\limits_0^s \left(- \dot R_{(2)(4)(2)(3)} -
\frac{M}{S_0}R_{(3)(4)(2)(3)}\right)
$$
$$ 
\times \sin\frac{M}{S_0}(s-x)\,dx + a_{(3)}(0) \cos\frac{M}{S_0}s 
$$
\begin{equation}\label{n+14}
 + \left(a_{(2)}(0) +\frac{S_0}{M}
R_{(2)(4)(2)(3)}(0)\right)\sin\frac{M}{S_0}s,
\end{equation}
where $a_{(2)}(0)$ and $a_{(3)}(0)$ are the corresponding  values at $s=0$. In the simple case of the Minkowski spacetime it follows from (\ref{n+13}) and (\ref{n+14} that 
\begin{equation}\label{n+15}
a_{(2)}(s) = a_{(2)}(0)\cos\frac{M}{S_0}s - a_{(3)}(0)\sin\frac{M}{S_0}s,
\end{equation}
\begin{equation}\label{n+16}
a_{(3)}(s) = a_{(3)}(0)\cos\frac{M}{S_0}s + a_{(3)}(0)\sin\frac{M}{S_0}s.
\end{equation}
As it can easily be seen, the expressions for $a_{(2)}(s)$ and $a_{(3)}(s)$ in (\ref{n+15}) and (\ref{n+16}) do not contain oscillatory terms if and only if $a_{(2)}(0)=0$ and $a_{(3)}(0)=0$. When one of the values $a_{(2)}(0)$ and $a_{(3)}(0)$ or both of them are nonzero, Eqs. describe the known oscillatory solutions.

It gets more interesting when we investigate some special solutions of the strict MP equations (\ref{1a}) and (\ref{2a}) under condition (\ref{3a}) in a spacetime with some nonzero curvature. Indeed, let us suppose that we obtained a partial solution of these equations in a certain background. Theoretically, after calculations of comoving tetrad components we can obtain explicit expressions of $R_{(2)(4)(2)(3)}$ and $R_{(3)(4)(2)(3)}$ as functions of $s$.Then by performing corresponding integration in the right-hand side of Eqs. (\ref{n+13}) and (\ref{n+14}) explicit expressions for $a_{(2)}(s)$ and $a_{(3)}(s)$ can be written. In Eqs. (\ref{n+13}) and (\ref{n+14}) $a_{(2)}(0)$ and $a_{(3)}(0)$ are the constants of integration, and different numerical values of  $a_{(2)}(0)$ and $a_{(3)}(0)$ correspond to different types of particle motions, both oscillatory and nonoscillatory. As a result, due to the obtained expressions for Eqs. (\ref{n+13}) and (\ref{n+14}) it is possible to identify the partial solution of the MP equations under condition (\ref{3a}) as the one which describes some helical or nonhelical motion. It is appropriate to illustrate this procedure with considered above highly relativistic circular orbits in Schwarzschild-de Sitter background. For this purpose we take into account that according to Eqs. (\ref{5}) and (\ref{6}) all components
$R_{(\pi)(\rho)(\sigma)(\tau)}$ are constant on circular orbits because in this case $u^1=0$, $u^3=const$, $u^4=const$, and $r=const$ (we also note that in this case  the relationship $\gamma_{(i)(k)(4)}=0$ is satisfied). Then the integrals in the right-hand side of Eqs. (\ref{n+13}) and (\ref{n+14}) can be calculated easily and we obtain
$$
a_{(2)}(s) = - R_{(2)(4)(2)(3)}\frac{S_0}{m}\left(1- \cos\frac{M}{S_0}s\right)
$$
\begin{equation}\label{n+17}
- R_{(3)(4)(2)(3)}\frac{S_0}{m}\sin\frac{M}{S_0}s +
a_{(2)}(0)\cos\frac{M}{S_0}s - a_{(3)}(0)\sin\frac{M}{S_0}s,
\end{equation}
$$
a_{(3)}(s) =  - R_{(3)(4)(2)(3)}\frac{S_0}{m}\left(1- \cos\frac{M}{S_0}s\right)
$$
\begin{equation}\label{n+18}
+ R_{(2)(4)(2)(3)}\frac{S_0}{m}\sin\frac{M}{S_0}s +
a_{(3)}(0)\cos\frac{M}{S_0}s + a_{(2)}(0)\sin\frac{M}{S_0}s.
\end{equation}
It follows from Eqs. (\ref{n+17}) and (\ref{n+18}) that the oscillatory terms in the right-hand sides of these equations can be excluded if and only if the values $a_{(2)}(0)$ and $a_{(3)}(0)$ are
\begin{equation}\label{n+19}
a_{(2)}(0) = -\frac{S_0}{m}R_{(2)(4)(2)(3)}, \quad
a_{(3)}(0) = -\frac{S_0}{m}R_{(3)(4)(2)(3)}. 
\end{equation}
Note that these values coincide with the constant values of $a_{(2)}$ and 
$a_{(3)}$ on the circular orbits which were considered above.

The behavior of a spinning particle in the Schwarzschild-de Sitter background was investigated in \cite{Stu, Mort, Kunst} without focusing just on specific cases of highly relativistic motions. In \cite{Stu} the stationary equilibrium condition for a spinning particle located in the Schwarzschild-de Sitter background was studied using the MP equations with relationship (\ref{3a}). This condition is
$$
\frac{M}{r} = \frac{\Lambda r^2}{3},
$$
and then the gravitational attraction is compensated by the cosmological repulsion.
The MP equations with condition (\ref{4a}) were studied in \cite{Mort} to find the location of the turning points when a spinning particle is moving in the equatorial plane of the Schwarzschild-de Sitter metric. The specific case with the isofrequency  pairing of spinning particles in the Schwarzschild-de Sitter background was investigated by the
MP equations under condition (\ref{4a}).

\section{Conclusions}

The above considered situations with highly relativistic motions of a spinning particle in the Schwarzschild-de Sitter background give the new theoretical data concerning physical effects following from general relativity. At the same time, it is useful to take into account the corresponding results in the practical high energy physics, astrophysics, and cosmology. 

For understanding the physical reason which determines the specific features of highly relativistic motions of a spinning particle in the Schwarzschild-de Sitter background it is important to keep in mind that in the proper system of reference such a particle feels the strong action of the gravitomagnetic field, as it is estimated in Secs. II and III. This action is repulsive or attractive depending on the correlation between the particle's spin and orbital velocity: the corresponding examples
are presented in 
Sec. IV for highly relativistic circular orbits.

\end{document}